\documentclass[aps,pra,twocolumn,amsmath,amssymb,nofootinbib,showpacs]{revtex4}
\usepackage[english]{babel}
\usepackage{latexsym}
\usepackage{graphics}
\usepackage{epsfig}
\usepackage{color}
\usepackage{bm}
\usepackage{amsmath}
\usepackage{amssymb}

\begin{document}

\title{Elastic Multi-Body Interactions on a Lattice}

\author{D.~S.~Petrov}
\affiliation{Universit\'e Paris-Sud, CNRS, LPTMS, UMR8626, Orsay, F-91405, France}

\date{\today}

\begin{abstract}

We show that by coupling two hyperfine states of an atom in an optical lattice one can independently control two-, three-, and four-body on-site interactions in a non-perturbative manner. In particular, under typical conditions of current experiments one can have a purely three- or four-body interacting gas of $^{39}$K atoms characterized by on-site interaction shifts of several 100Hz.

\end{abstract}


\maketitle

Effective multi-body interactions can arise even in a purely two-body interacting system when one integrates out some of its (high energy) degrees of freedom or substitutes the actual two-body potential by a pseudopotential. Such effective forces are important in many fields from nuclear and high-energy physics to ultracold gases. A textbook example is the appearance of an effective three-body force in the zero-range pseudopotential description of the hard sphere Bose gas \cite{LHYPR1957,WuPR1959}. Inclusion of a three-body renormalizing potential (three-body parameter) is unavoidable for a resonantly interacting Bose gas \cite{HammerRMP2013}. We can name a few other systems for which multi-body interactions are important. The quasi-1D Bose gas is kinematically 1D, but virtual transversal excitations lead to the appearance of local three-body terms in the corresponding effective 1D model and break integrability \cite{Muryshev2002,Sinha,Mazets2008}. Similar terms enter the single Landau level description of quantum Hall systems when one takes into account virtual excitations to other Landau levels \cite{BisharaNayak2009,PetersonNayak2013,SodemannMacDonald2013}. Inclusion of these small corrections can lead to qualitative modifications of the phase diagram and can stabilize phases otherwise predicted to be unstable. For bosonic atoms in an optical lattice, effective multi-body interactions emerge when one reduces this continuum system to the single-band Hubbard model \cite{LiNJP2006,JohnsonNJP2009,HazzardPRA2010,TiesingaPRA2011}. In spite of their weakness compared to the two-body interaction, they can be measured spectroscopically \cite{CampbellScience2006} and give rise to a peculiar collapse and revival dynamics \cite{GreinerNature2002,AnderliniJPB2006,SebbyPRL2007,WillNature2010}.

In recent years various possibilities to independently control multi-body interactions have been discussed and a number of more or less technically complicated schemes has been suggested \cite{Mehen,CooperPRL2004,Buchler,DaleyZoller,Roncaglia,Mazza,Tiesinga,KapitSimon,Taylor,Daley,PetrovPRL2014}. The task is not straightforward but highly rewarding because of many potentially interesting implications, in particular, for the creation of topological quantum Hall phases \cite{NayakRMP2008}, stabilizing paired bosonic superfluids \cite{Radzihovsky,Stoof,Lee,Mueller,DaleyZoller}, observing self-trapped droplets \cite{Bulgac} and other phenomena \cite{Gammal,ParedesPRA2007,ChenPRA2008,SchmidtPRL2008,CapogrossoPRB2009,DiehlPRB2010,NgPRB2011,BonnesPRL2011}. 

A system in which the (effective) three-body interaction is finite and the (effective) two-body one is negligible is unnatural but not impossible. It requires that the two-body on-shell scattering amplitude vanish but the off-shell one remain large. The latter ensures a significant distortion of the wavefunction and leads to higher-order multi-body effects. Such an off-shell interaction can be realized by using a sufficiently exotic interaction potential or single-particle wave functions. A good starting point is to consider a finite potential which averages to zero (vanishes in the first order Born approximation) and is then modified in such a way that the higher order attraction is also compensated. In Ref.~\cite{PetrovPRL2014} we have considered the interlayer potential for dipoles in the bilayer geometry (which averages to zero) and have shown a way to tune it to a two-body zero crossing, at the same time obtaining a strong three-body repulsion.

In this paper we extend this idea to a two-component Bose gas in an optical lattice. In this case one can make single-particle wavefunctions {\it exotic} by coupling two internal states with a nearly resonant field (the so-called free-free transition \cite{Hanna2010}). By varying the corresponding Rabi frequency $\Omega$ and detuning $\Delta$ one can rotate the wavefunction in the space of the two dressed states and thus tune the two-body interaction, for instance, to a zero crossing. We show that in contrast to what one can obtain near a usual Feshbach zero crossing \cite{FattoriPRL2008,PollackPRL2009,Shotan2014}, in our case multi-body interactions can be made much stronger and elastic. Curiously, the same technique without additional efforts can be used to make the two- and three-body interactions vanish while keeping a finite four-body one. We discuss implications of these results for current experiments and show that favorable conditions (suitable window of inter- and intrastate scattering lengths) are provided by hyperfine states $F=1,m_F=0$ and $F=1,m_F=-1$ of $^{39}$K.

Considering the frequency of the hyperfine transition (typically 10$^7$-10$^8$Hz) the largest energy scale in our problem we write the Hamiltonian in the rotating wave approximation as
\begin{eqnarray}\label{Ham}
H&=& \int_{\pmb{r}}\left\{\sum_{\sigma} \Psi^\dagger_{\sigma {\boldsymbol{r}}} [-\nabla^2_{\boldsymbol{r}}/2+V_{\rm ext}({\boldsymbol{r}})]\Psi_{\sigma {\boldsymbol{r}}}\right.\nonumber\\ 
&&\hspace{-1.cm}+\left.\frac{\Delta}{2} (\Psi^\dagger_{\downarrow {\boldsymbol{r}}}\Psi_{\downarrow {\boldsymbol{r}}}-\Psi^\dagger_{\uparrow {\boldsymbol{r}}}\Psi_{\uparrow {\boldsymbol{r}}})
 -\frac{\Omega}{2} (\Psi^\dagger_{\uparrow {\boldsymbol{r}}}\Psi_{\downarrow {\boldsymbol{r}}}+\Psi^\dagger_{\downarrow {\boldsymbol{r}}}\Psi_{\uparrow {\boldsymbol{r}}})\right\}\nonumber\\
&&\hspace{-1.cm}+\frac{1}{2} \int_{{\boldsymbol{r}},{\boldsymbol{r}}'}\sum_{\sigma,\sigma'} \Psi^\dagger_{\sigma {\boldsymbol{r}}} \Psi^\dagger_{\sigma' {\boldsymbol{r}}'}V_{\sigma\sigma'}(|{\boldsymbol{r}}-{\boldsymbol{r}}'|)\Psi_{\sigma {\boldsymbol{r}}} \Psi_{\sigma' {\boldsymbol{r}}'},
\end{eqnarray} 
where $\Psi^\dagger_{\sigma {\boldsymbol{r}}}$ is the creation operator of a boson in the internal (dressed) state $\sigma(=\uparrow,\downarrow)$ with coordinate ${\boldsymbol{r}}$, $V_{\rm ext}$ is the external potential of the optical lattice, and $V_{\sigma\sigma'}(r)$ are the short-range interparticle interactions, which are characterized by the $s$-wave scattering lengths $a_{\sigma\sigma'}$, and we adopt the units $\hbar=m=1$. 

For a single particle, the orbital and spinor degrees of freedom, respectively described by the first and second lines in Eq.~(\ref{Ham}), decouple. The former is characterized by the usual band structure in the periodic potential $V_{\rm ext}$ and the diagonalization of the latter gives two spinor eigenstates split in energy by $\sqrt{\Omega^2+\Delta^2}$. 
We will assume that the temperature of the system is lower than this spinor gap so that the gas is effectively spinless. However, this gap should not be too large in order to allow for virtual excitations of the upper spinor branch during collisions. The lower the gap, the stronger are the off-shell effects and multi-body interactions. 

We will further assume that $\Omega$ and the {\it bare} on-site interaction shifts $g_{\sigma\sigma'}$ are (i) much smaller than the on-site confinement frequencies and (ii) much larger than the intersite tunnelling amplitude $t$. Condition (i) allows us to use the single orbital mode approximation and completely neglect virtual excitations to higher orbital bands considering the spinor sector as the major source of effective interactions. Assumption (ii) ensures that when a particle tunnels, the wavefunction has enough time to adjust itself to the ground state for the new configuration of the on-site occupations. It also allows us to neglect the nearest neighbor and more distant effective interactions \cite{RemLarget}.

With these assumptions we reduce our original problem to the spinless Bose-Hubbard model with the on-site energy term
\begin{equation}\label{OnSiteEn}
E(N)=-\frac{\sqrt{\Omega^2+\Delta^2}}{2}N+\sum_{i=2}^{N}U_i\frac{N!}{i!(N-i)!},
\end{equation}
which is found by diagonalizing the (particle number conserving) on-site Hamiltonian
\begin{equation}\label{OnSiteHam}
H_0=\frac{\Delta}{2} (b^\dagger_\downarrow b_\downarrow-b^\dagger_\uparrow b_\uparrow)
 -\frac{\Omega}{2} (b^\dagger_\uparrow b_\downarrow +b^\dagger_\downarrow b_\uparrow)+\sum_{\sigma,\sigma'} \frac{g_{\sigma\sigma'}}{2} b^\dagger_\sigma b^\dagger_{\sigma'} b_\sigma b_{\sigma'}.
\end{equation} 
Using the harmonic approximation for the on-site confinement, the interaction constants equal $g_{\sigma\sigma'}=\sqrt{2/\pi}a_{\sigma\sigma'}/l_xl_yl_z$, where $l_x$, $l_y$, and $l_z$ are the oscillator lengths. 

For a given $N$ one can use the set of $N+1$ wave functions $|i,N-i\rangle$ describing the Fock states of $i$ $\uparrow$ bosons and $N-i$ $\downarrow$ ones. The Hamiltonian (\ref{OnSiteHam}) in this representation becomes a symmetric tridiagonal matrix with diagonal elements 
\begin{align}\label{MatDiag}
&\langle i,N-i|H_0|i,N-i\rangle =  \Delta(N-2i)/2+g_{\uparrow\uparrow}i(i-1)/2\nonumber\\
&+ g_{\uparrow\downarrow}i(N-i)+g_{\downarrow\downarrow}(N-i)(N-i-1)/2
\end{align}
and off-diagonal ones
\begin{equation}\label{MatOffDiag}
\langle i,N-i|H_0|i+1,N-i-1\rangle = -\frac{\Omega\sqrt{(N-i)(i+1)}}{2}.
\end{equation} 
Its diagonalization is straightforward and Eq.~(\ref{OnSiteEn}) can be applied iteratively to find $U_N$, given the knowledge of $U_M$ for all $M<N$. 

The five-dimensional parameter space $\{\Omega, \Delta, g_{\sigma\sigma'}\}$ provides enough freedom for an independent control over $U_N$, at least for several lowest $N$. When $\Delta$, $\Omega$, and $g_{\sigma\sigma'}$ are of the same order of magnitude the problem is non-perturbative, the multi-body interaction constants $U_N$ are comparable to each other (cf. \cite{JohnsonNJP2009,WillNature2010}), and quite exotic combinations of them are possible. However, let us limit our discussion to the most radical {\it $N$-body interacting} case, in which finite $U_N$ comes along with vanishing (or very small) $U_M$ for all $M<N$. First we discuss the three-body interacting case taking into account, as much as possible, current experimental constraints. This sets the following optimization problem. For a given combination of $g_{\sigma\sigma'}$ maximize $U_3>0$ with respect to $\Omega$ and $\Delta$ with the constraint $U_2=0$.

Most clearly the mechanism behind the effective three-body interaction can be seen for $\Omega=\Delta=g_{\uparrow\downarrow}=0$ and $g_{\downarrow\downarrow}=g_{\uparrow\uparrow}=g>0$. In this case the two-body ground state is $|1,1\rangle$ leading to $U_2=0$. The ground state for $N=3$ is doubly degenerate, spanned by $|2,1\rangle$ and $|1,2\rangle$. The effective three-body interaction equals $U_3=g$ and is generated by the spinor frustration: each pair prefers to be in the $\uparrow\downarrow$ singlet state -- the condition, which can not be simultaneously satisfied for all $N>2$ particles. It is thus crucial that there is only two internal states. Vanishing $\Omega$ is not consistent with some of our initial assumptions, but it is clear that the result does not change much if $t,T\ll \Omega\ll g$, we still arrive at $U_3\approx g$.

In the case of generally different $g_{\sigma\sigma'}$ the solution of our optimization problem is $\Omega=0$, $\Delta=g_{\uparrow\downarrow}{\rm sign}(g_{\downarrow\downarrow}-g_{\uparrow\uparrow})$, the maximum equals
\begin{equation}\label{U3} 
U_{3,{\rm max}}=\begin{cases}
{\rm min}(g_{\downarrow\downarrow},g_{\uparrow\uparrow}), &|g_{\downarrow\downarrow}-g_{\uparrow\uparrow}|>-g_{\uparrow\downarrow},\\
{\rm max}(g_{\downarrow\downarrow},g_{\uparrow\uparrow})+g_{\uparrow\downarrow}, &|g_{\downarrow\downarrow}-g_{\uparrow\uparrow}|<-g_{\uparrow\downarrow},
\end{cases}
\end{equation} 
and the inequalities 
\begin{equation}\label{TheInequalities}
0<{\rm min}(g_{\downarrow\downarrow},g_{\uparrow\uparrow}),\; -{\rm max}(g_{\downarrow\downarrow},g_{\uparrow\uparrow})<g_{\uparrow\downarrow}<0
\end{equation}
define the interesting for us region where $U_2=0$ and $U_3>0$. Indeed, by treating the off-diagonal terms of $H_0$ perturbatively, one can show that for any given combination of $g_{\sigma\sigma'}$ satisfying (\ref{TheInequalities}), the point $\Omega=0$, $\Delta=g_{\uparrow\downarrow}{\rm sign}(g_{\downarrow\downarrow}-g_{\uparrow\uparrow})$ is a (local) maximum of $U_3$ along the curve $U_2(\Omega,\Delta)=0$. The correction is quadratic in $\Omega$ and one can introduce a finite $\Omega$ while maintaining $U_3$ close to this maximum. Note that the three-body interaction obtained in this manner is linear in $g_{\sigma\sigma'}$. This result is to be compared with the dependence $U_3\propto g^2/\omega$ which arises from virtual excitations to higher {\it orbital} bands with the interband spacing given by the on-site oscillation frequency $\omega$ \cite{JohnsonNJP2009}. In our case the quadratic dependence $U_3\propto g^2/\Omega$ would arise for $g_{\sigma\sigma'}\ll \Omega$ in the second order perturbation theory. Thus, the gain in the amplitude of the three-body effective interaction in the spinor case compared to the orbital one is due to a smaller gap ($\Omega\ll\omega$) between the low-energy and high-energy (virtual) degrees of freedom. Accordingly, $U_3$ is maximized in the most non-perturbative limit $\Omega\rightarrow 0$.

We apply the above formalism to the case of $^{39}$K in which $a_{\sigma\sigma'}$ for various hyperfine states have been studied theoretically \cite{DErricoNJP2007,Lysebo} and experimentally \cite{DErricoNJP2007}. In particular, conditions (\ref{TheInequalities}) are satisfied for the second and third lowest hyperfine states, $F=1,m_F=0$ ($\sigma=\downarrow$) and $F=1,m_F=-1$ ($\sigma=\uparrow$), in the magnetic field region from $B=56$ to 59 G. More specifically, in this region $a_{\uparrow\uparrow}$ decreases from approximately 1.85 to 1.56nm, $a_{\uparrow\downarrow}$ increases from -2.83 to -2.75nm, and the point $B_0=59.3(6)$ marks a Feshbach resonance in the $\downarrow\downarrow$ channel with the width $\Delta B\approx -10$G and background scattering length of approximately -0.95nm. 

For concreteness let us choose $a_{\downarrow\downarrow}= 9.4$nm, $a_{\uparrow\uparrow}= 1.7$nm, and $a_{\uparrow\downarrow}= -2.8$nm, which should be, within the claimed theoretical and experimental errorbars, a good estimate of the scattering lengths at about -1G detuning from the resonance. Then, let us assume an optical lattice with the lattice constant $\lambda/2=532$nm and intensity $V_0=15E_R$ [$E_R=2\pi^2\hbar^2/m\lambda^2$] in each of the three spatial directions. This produces \cite{BlochRMP2008} a three-dimensional lattice with an isotropic on-site confinement of the frequency $\omega\approx 2\pi\times 35$kHz (oscillator lengths $l=l_x=l_y=l_z\approx 86$nm), intersite tunneling amplitude $t\approx 2\pi\times 30$Hz, and the on-site interaction shifts $g_{\downarrow\downarrow}\approx 2\pi\times 3.05$kHz, $g_{\uparrow\uparrow}\approx 2\pi\times 0.55$kHz, and $g_{\uparrow\downarrow}\approx -2\pi\times 0.91$kHz. 

\begin{figure}
\centerline{\includegraphics[width=0.9\hsize,clip,angle=0]{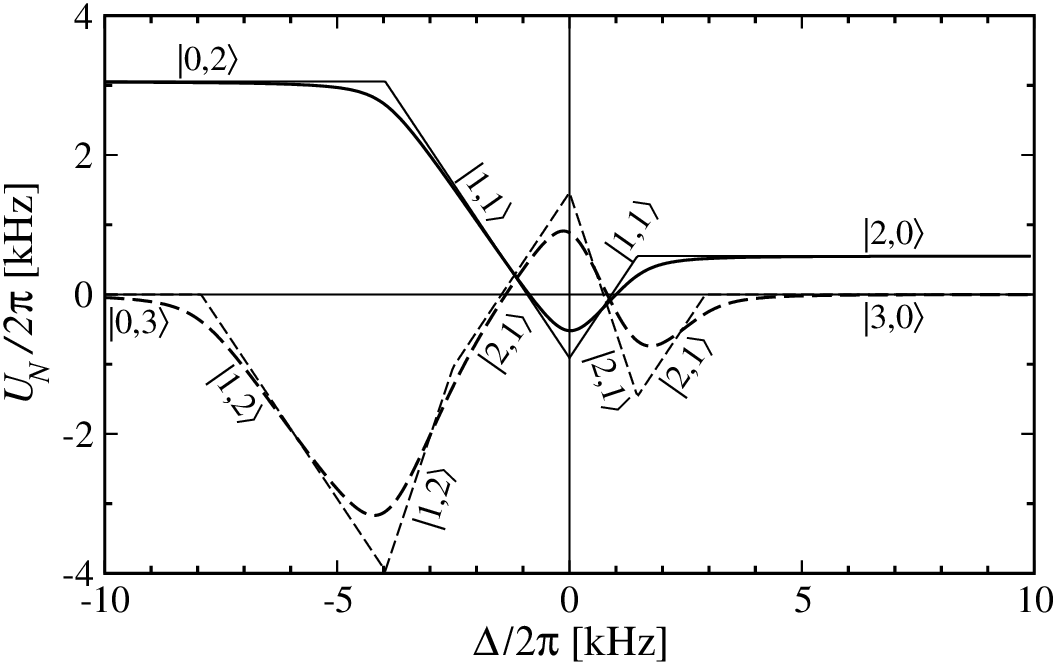}}
\caption{$U_2$ (thick solid) and $U_3$ (thick dashed) versus $\Delta$ for $\Omega=2\pi\times 0.5$kHz. The thin piecewise linear curves correspond to the limit of vanishing $\Omega$ where Fock states $|i,N-i\rangle$ are exact eigenstates (see labels). The interesting for us point where $U_2=0$ and $U_3>0$ is given by $\Delta=-2\pi\times 0.87$kHz, $U_3= 2\pi\times 0.48$kHz and can be compared to the case $\Omega=0$ where $\Delta=g_{\uparrow\downarrow}=-2\pi\times 0.9$kHz, $U_3=g_{\uparrow\uparrow}=2\pi\times 0.55$kHz.}
\label{fig:U2and3general}
\end{figure}

In Fig.~\ref{fig:U2and3general} we plot $U_2$ (thick solid line) and $U_3$ (thick dashed line) versus $\Delta$ for $\Omega=2\pi\times 0.5$kHz. For comparison we also show the case $\Omega\rightarrow 0$ where the ground states are Fock states and $U_2(\Delta)$ and $U_3(\Delta)$ become piecewise linear functions. Each segment of them is labeled accordingly and the corresponding values of $E(N)$ and $U_N$ can be restored from Eqs.~(\ref{OnSiteEn}) and (\ref{MatDiag}). For small finite $\Omega$ the segment junctions smoothen and follow lower or upper branches of 3-body, 2-body, or 1-body (for $\Delta=0$) level anticrossings. In the shown example of $\Omega=2\pi\times 0.5$kHz there are two zero crossings of $U_2$. The right one corresponds to negative $U_3$, but at the left crossing point we obtain $U_3=2\pi\times 0.48$kHz, to be compared with $U_{3,{\rm max}}=g_{\uparrow\uparrow}=2\pi\times 0.55$kHz [see Eq.~(\ref{U3})]. We see that a rather strong elastic three-body effective interaction can coexist with the vanishing two-body one. 

A possible practical issue related to this proposal is that three atoms on a single site can recombine to a deeply bound molecule. This loss process can be accounted for by a negative imaginary part of $U_3$, which, for non-resonant two-body interactions, is proportional to $(R_{vdW}^4/m)\int|\phi_0({\bf r})|^6d^3r \propto R_{\rm vdW}^4/ml^6$. Here $\phi_0$ is the on-site ground state wavefunction and the van der Waals range $R_{\rm vdW}$ is of the same order of magnitude as $a_{\sigma\sigma'}$. Note that $|{\rm Im}U_3|/{\rm Re}U_3\sim (R_{\rm vdW}/l)^3$ is very small. More quantitatively, by adopting the free space loss rate formula to the case of a single confined triple we derive ${\rm Im}U_3=-(K_3/3!)\int|\phi_0({\bf r})|^6d^3r$, where $K_3$ is the three-body recombination loss rate constant for non-condensed atoms. For non-resonant $^{39}$K it is rather small \cite{ZaccantiNatPhys2009}, $K_3<10^{-29}$cm$^6$/s. For the considered scattering lengths, to be on the safe side, we assume $K_3<10^{-27}$cm$^6$/s and arrive at $-{\rm Im}U_3< 2\pi\times 0.4$Hz$\ll {\rm Re}U_3$.

\begin{figure}
\centerline{\includegraphics[width=0.9\hsize,clip,angle=0]{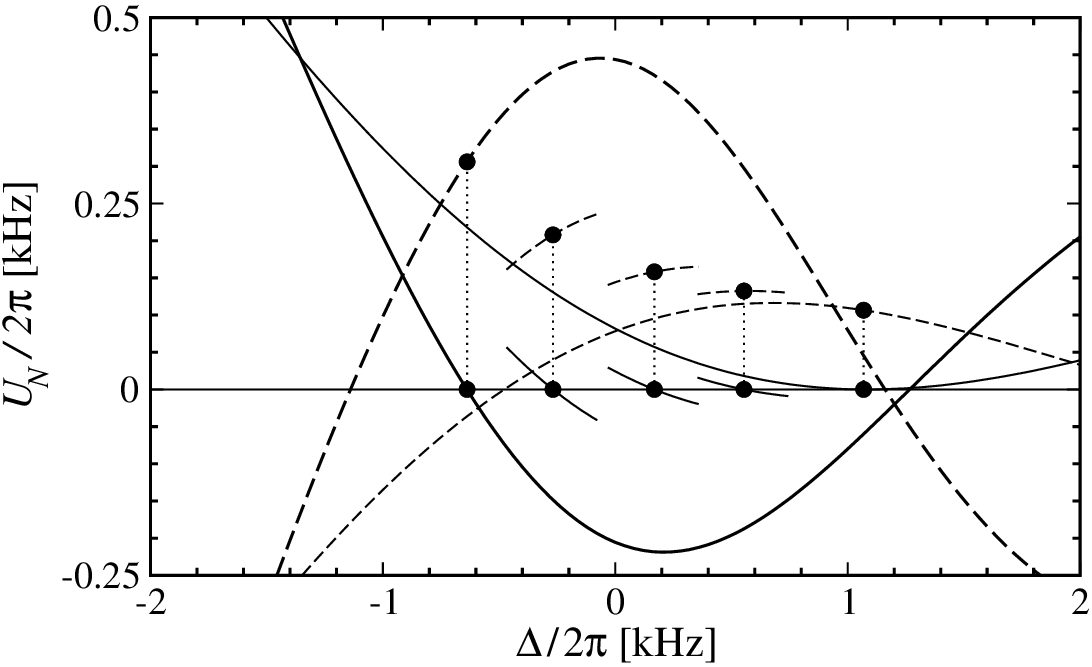}}
\caption{$U_2$ (solid) and $U_3$ (dashed) versus $\Delta$ for $\Omega/2\pi= 1.3$, $2.06$, $2.69$, $3.09$, and $3.33$kHz in the vicinities of two-body zero crossings (circles). The whole curves are plotted for the lowest (thick) and highest (thin) values of $\Omega$. At the crossings $-dU_2/d\Delta=1/2$, 1/4, 1/8, 1/16, and 0, respectively.}
\label{fig:U2and3zoom}
\end{figure}

Another potential problem can be fluctuations $\delta B$ of the magnetic field causing an instability of the resonant radio frequency $\Delta_0(B)$ for the hyperfine transition, which, in turn, gives rise to fluctuations of the two-body interaction $\delta U_2=-(dU_2/d\Delta)(d\Delta_0(B)/dB)\delta B$. The first derivative $d\Delta_0(B)/dB$ depends on the atom, hyperfine states, and magnetic field. For the considered hyperfine states of $^{39}$K at $B\approx 58$G it equals $d\Delta_0(B)/dB\approx 2\pi\times 0.7$kHz/mG \cite{Rem39K}. The derivative $dU_2/d\Delta$ for small $\Omega$ equals $dU_2/d\Delta\approx -1$. In this case, in order to realize $|U_3|\gg |U_2|$, the magnetic field fluctuations should be kept below 1mG. However, $dU_2/d\Delta$ decreases with $\Omega$ faster than $U_3$. In Fig.~\ref{fig:U2and3zoom} we show $U_2(\Delta)$ and $U_3(\Delta)$ close to two-body zero crossings for $\Omega$ chosen such that at the crossings $-dU_2/d\Delta=1/2$, 1/4, 1/8, 1/16, and 0, respectively. To avoid cluttering we show the whole curves only for the cases $\Omega=2\pi\times 1.3$kHz ($dU_2/d\Delta=-1/2$) and $\Omega=2\pi\times 3.33$kHz for which $U_2(\Delta)$ touches the horizontal axis ($dU_2/d\Delta=0$). In the latter case $U_3$ is just above 100Hz, but the restriction on the magnetic field stability is relaxed.

\begin{figure}
\centerline{\includegraphics[width=0.9\hsize,clip,angle=0]{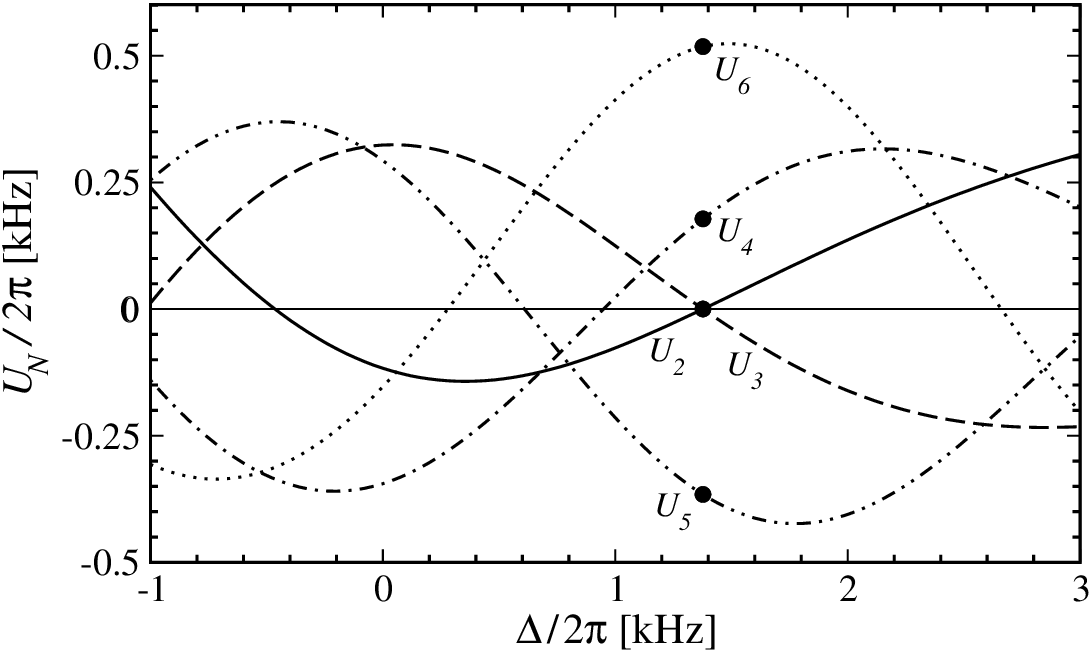}}
\caption{$U_2$ (solid), $U_3$ (dashed), $U_4$ (dash-dotted), $U_5$ (dash-doubledotted), and $U_6$ (dotted) versus $\Delta$ for $\Omega/2\pi= 1.7$kHz. The circles indicate the corresponding values at the simultaneous two- and three-body zero crossing.}
\label{fig:U4and5}
\end{figure}

Let us now turn to the four-body and higher order interactions. The $N$-body interacting case, $U_2=U_3=...=U_{N-1}=0$ and $U_N>0$, can in principle be realized by extending the spin frustration idea to an atom with $N-1$ internal states, provided repulsive intrastate interactions and attractive interstate ones. Then, $N-1$ atoms on a single site can avoid the intrastate repulsion by occupying different internal states. However, for larger number of atoms at least two of them have to be in the same state leading to a positive energy shift.

It turns out that the four-body interacting case can be realized by coupling only two internal states. Indeed, in Fig.~\ref{fig:U2and3zoom} we notice that the points $U_2=U_3$ for $\Omega=2\pi\times 1.3$kHz and for $\Omega=2\pi\times 3.3$kHz are on different sides of the horizontal axis. We find that $U_2=U_3=0$ for $\Omega=2\pi\times 1.7$kHz, $\Delta=2\pi\times 1.38$kHz. At this point the four-body interaction is repulsive and equals $U_4=2\pi\times 0.18$kHz (see Fig.~\ref{fig:U4and5}). We have checked that $U_4$ can be increased (keeping $U_2=U_3=0$) by decreasing the magnetic field detuning from the $\downarrow\downarrow$ Feshbach resonance (at -0.5G detuning we obtain $U_4\approx 0.33$kHz). However, close to the resonance $a_{\downarrow\downarrow}$ becomes comparable to the oscillator length of the on-site confinement and we can no longer rely on the single-mode approximation. We also expect ${\rm Im}U_3$ to increase. Nevertheless, our findings indicate that the four-body interacting case, which seems to be too exotic, is reachable in current experiments.

In Fig.~\ref{fig:U4and5} we also show $U_5$ and $U_6$ for reference. Their absolute values are actually larger than $U_4$ emphasizing the non-perturbative nature of the problem. However, the largest contribution to $E(N)$ typically comes from the lowest non-vanishing $U(M)$. At the crossing in the conditions of Fig.~\ref{fig:U4and5} $U_5$ is negative and equals $-2\pi\times 0.366$kHz, which is twice as large as $U_4$. Yet, the on-site interaction energy of five atoms equals $5U_4+U_5>0$.

Finally, let us discuss possible experimental signatures of the few-body interactions. The powerful method of Ref.~\cite{WillNature2010} is capable of accurately resolving even very weak multi-body interactions. However, our interactions are non-perturbative and we expect much stronger effects and qualitative changes of the many-body phase diagram. For example, a superfluid with $U_2<0$ and $U_3>0$ (both weaker or comparable to $t$) should be in the droplet state \cite{Bulgac}. In the absence of external trapping (the optical lattice is kept) it would exhibit a soliton-like self-trapping with a flat density $n=-3U_2/2U_3$. Then, increasing both $U_2$ and $U_3$ should eventually lead to the paired state \cite{DaleyZoller,DiehlPRB2010,NgPRB2011,BonnesPRL2011}. Another manifestation of multi-body interactions is the modification of the Mott-superfluid lobes \cite{ChenPRA2008}. In particular, sufficiently deeply in the Mott-insulating state with $n=2$ atoms per site the excitation gap equals $E(3)+E(1)-2E(2)=U_3+U_2$. From Fig.~\ref{fig:U2and3general} we see that this gap decreases and eventually vanishes as we go from large negative $\Delta$ to the point $\Delta\approx -2\pi\times 5$kHz, which should be detectable by standard methods \cite{GreinerNature2002n2,BlochRMP2008}. Note that the insulating state with $n=1$ does not feel $U_3$ and thus stays incompressible.

The local $N$-body repulsive interaction is the parent Hamiltonian for the $N-1$st member of the Read-Rezayi series of quantum Hall states \cite{ReadPRB1999}. The second, known as the Moore-Read state, and particularly the third one (Read-Rezayi state) are seriously considered for the universal topological quantum computing \cite{NayakRMP2008}. The proposal heavily relies on the gap ($\propto U_N$) which protects the quantum state against local perturbations. Curious to estimate the gap for the Read-Rezayi state in the ``parent'' four-body interacting case we have performed exact diagonalization for 12 bosons with 8 vortices (effective magnetic fluxes) on a rectangular lattice with periodic boundary conditions. For lattices larger than $5\times 8$ sites (number of fluxes per plaquette $<1/5$ and $n<3/10$), as expected, we observe four almost degenerate almost zero energy ground states well separated from the others by the gap $\approx 0.036 U_4 n^3$ (quite close to the naive estimate $U_4n^3/4!$). It is small for these densities but for larger $n$ we face finite size effects and the Hofstadter broadening of the Landau level \cite{Hofstadter,KapitPRL2010}. A more systematic study for larger systems is desirable in order to identify combinations of $U_N$ which maximize the gap for these interesting topological phases. We should point out that some proposals for creating artificial gauge potentials for neutral atoms \cite{Dalibard} also rely on using the hyperfine structure and can thus interfere with our way of generating multi-body interactions. Nevertheless, our method should commute with shaking and/or rotating the lattice, and with other schemes (see \cite{CooperAdvPhys2008,Dalibard} for review).

We thank N. Cooper and M. Zaccanti for fruitful discussions and A. Simoni and M. Lysebo for providing the data of Refs.~\cite{DErricoNJP2007,Lysebo} in numerical form, and acknowledge support from the IFRAF Institute and the Institute for Nuclear Theory during the program {\it Universality in Few-Body Systems: Theoretical Challenges and New Directions, INT-14-1}. The research leading to these results has received funding from the European Research Council under European Community's Seventh Framework Programme (FR7/2007-2013 Grant Agreement no.341197).

\end{document}